\documentclass[twocolumn,showpacs,preprintnumbers,amsmath,amssymb, superscriptaddress]{revtex4}

\usepackage{amsmath}
\usepackage{amsfonts}   
\usepackage{graphics}  
\usepackage{psfrag} 
\usepackage{graphicx} 
\usepackage{epsfig}    
\usepackage{rotating}  
 \usepackage[latin1]{inputenc}
 \usepackage{color}
 \usepackage{rotating}

\usepackage{color}
\usepackage{epsfig}

\definecolor{darkgreen}{rgb}{0,0.5,0} 
\definecolor{violet}{rgb}{0.5,0,0.5}
\definecolor{orange}{rgb}{0.2,0.5,0.5}

\newcommand\scalemath[2]{\scalebox{#1}{\mbox{\ensuremath{\displaystyle #2}}}}

\begin{document}
\preprint{}

\title{Mobility, fitness collection, and the breakdown of cooperation}

\author{Anatolij Gelimson}
\affiliation{Arnold Sommerfeld Center for Theoretical Physics and Center for NanoScience, Department of Physics, Ludwig-Maximilians-Universit\"at M\"unchen, Theresienstr. 37, D-80333 M\"unchen, Germany}
\affiliation{The Rudolf Peierls Centre for Theoretical Physics, University of Oxford, 1 Keble Road, Oxford, OX1 3NP, United Kingdom}

\author{Jonas Cremer}
\affiliation{Arnold Sommerfeld Center for Theoretical Physics and Center for NanoScience, Department of Physics, Ludwig-Maximilians-Universit\"at M\"unchen, Theresienstr. 37, D-80333 M\"unchen, Germany}
\affiliation{Center for Theoretical Biological Physics, University of California at San Diego, La Jolla, California 92093, USA}

\author{Erwin Frey}
\affiliation{Arnold Sommerfeld Center for Theoretical Physics and Center for NanoScience, Department of Physics, Ludwig-Maximilians-Universit\"at M\"unchen, Theresienstr. 37, D-80333 M\"unchen, Germany}
\begin{abstract} 

The spatial arrangement of individuals is thought to overcome the dilemma of cooperation: When cooperators engage in  clusters they might share the benefit of cooperation while being more protected against non-cooperating individuals, which benefit from cooperation but save the cost of cooperation. This is paradigmatically shown by the spatial prisoner's dilemma model. Here, we study this model in one and two spatial dimensions, but explicitly take into account that in biological setups fitness collection and selection are separated processes occurring mostly on vastly different time scales. This separation is particularly important to understand the impact of mobility on the evolution of cooperation. We find that even small diffusive mobility strongly restricts cooperation since it enables non-cooperative individuals to invade cooperative clusters. Thus, in most biological scenarios, where the mobility of competing individuals is an irrefutable fact, the spatial prisoner's dilemma alone cannot explain stable cooperation but additional mechanisms are necessary for spatial structure to promote the evolution of cooperation. The breakdown of cooperation is analyzed in detail. We confirm the existence of a phase transition, here controlled by mobility and costs, which distinguishes between purely cooperative and non-cooperative absorbing states. While in one dimension the model is in the class of the Voter Model, it belongs to the Directed Percolation (DP) universality class in two dimensions.
\end{abstract}


\maketitle

\section{Introduction}
While cooperative behavior is ubiquitous in nature, it is puzzling to understand from an evolutionary perspective: \emph{free-riders} benefit from cooperation but save the costs to provide a public good and hence have a fitness advantage compared to \emph{cooperators}~\cite{axelrod-1981-211,Maynard}.  To overcome this \emph{dilemma of cooperation}, additional mechanisms must be effective which select for cooperators. One important mechanism is the spatial clustering of cooperators where cooperators preferentially interact with other cooperators~\cite{Hamilton:1964,NowakSpatial,Roca:2009a, Durrett:1994}. This is illustrated by the \emph{spatial prisoner's dilemma}~\cite{NowakSpatial,Nowak:1994} where, in contrast to its famous well-mixed variant~\cite{axelrod-1981-211,Maynard}, individuals are arranged on a lattice and only interact with their nearest neighbors. Hence fitness explicitly depends on the composition of the neighborhood. Many different studies, employing a variety of deterministic and stochastic interaction rules, have confirmed that the formation of cooperative clusters can promote cooperation in such setups, see e.g. Refs.~\cite{Nakamuru:1997, Langer:2008, Szabo:2009, Helbing:2009, Fu:2010, Szolnoki:2009, Liu:2012, Szolnoki:2009a, AlonsoSanz:2009, Wang:2012}. This was also found for complex spatial structures like networks \cite{Ohtsuki:2006, vanBaalen:1998, Ifti:2004, Santos:2005, Abramson:2001, Lieberman:2005, Szabo, Vukov:2006}. Spatial clustering has also been studied experimentally. For example, Le Gac et al.~\cite{LeGac:2009} have found in experiments with toxin-producing \emph{E.~coli} that cooperation can also be maintained through cluster formation in a viscous environment and the degree of diffusion of the public good can be crucial for the evolutionary outcome. Theoretical studies also show that there can be a critical phase-transition into a purely cooperative state if costs fall below a certain threshold value~\cite{Szabo:1998, Szab'o:2002, Szabo:2005}.

Importantly, however, in spatial setups mobility can strongly challenge the evolution of cooperation~\cite{Hamilton:1964}. This is in agreement with recent studies by Chiong and Kirley~\cite{Chiong:2012}, who found a statistically not significant enhancement of cooperation for small mobility, and a strong reduction for intermediate mobility. For the spatial prisoner's dilemma, complex forms of mobility relying on more sophisticated abilities, like success driven mobility, have been shown to promote cooperation~\cite{Enquist:1993,Vainstein:2007,Sicardi:2009,Helbing:2008}. However, undirected mobility  (i.e.~diffusion), ubiquitous in biological situations, fosters the invasion of free-riders into cooperative clusters and thereby strongly counteracts the evolution of cooperation. Thus if undirected mobility threatens cooperation and mobility is undeniably a part of biological reality, this raises the question how fundamental spatial clustering as a mechanism to explain cooperation really is.

In this manuscript we investigate the impact of diffusive mobility on the level of cooperation and analyze the critical mobility where cooperation breaks down. To study this question, an understanding of the biological origin of fitness is crucial. While fitter individuals are selected for over many generations, fitness itself is the result of a multitude of different more microscopic processes like nutrient uptake or metabolic processes, which occur on much shorter time scales than reproduction. For example, when iron is lacking in the environment, cooperative strains of the proteobacteria \emph{Pseudomonas aeruginosa} produce iron-scavenging molecules (siderophores)~\cite{Diggle}. Released into the environment these molecules can efficiently bind single iron atoms and form complexes that can then be taken up by surrounding bacteria. Associated with the metabolic costs and benefits, the fitness of an individual changes with every siderophore production and iron-uptake process. However, reproduction and selection of fitter individuals only occurs after many such processes, on a longer time scale. This biologically more realistic origin of fitness has been mostly neglected in previous models by regarding fitness collection and reproduction as simultaneous events. In a population where building up fitness is a life-long dynamic process (\emph{fitness collection}), the individuals' mobility plays a particularly crucial role as the individuals' fitness strongly depends on the neighborhoods which they inhabit during their life span.

We here study a spatial prisoner's dilemma game and explicitly take fitness collection dynamics into account. With this formulation we are able to reproduce the expected limits. For vanishing mobility, we observe clustering and the typical spatial prisoner's dilemma dynamics.  In contrast, for high mobility, we recover the dynamics of the replicator equation for the well-mixed prisoner's dilemma. The transition between these scenarios is not smooth but separated by a critical non-equilibrium phase transition: Below critical costs and mobility values only cooperators remain in the long run while above only free-riders persist. 

\section{Model Definition}
Consider $N^d$ individuals occupying the sites of a regular $d$-dimensional square lattice with a linear extension $N$ and periodic boundary conditions. Each individual is either a cooperator ($C$) or free-rider ($F$) and individuals only interact with their nearest neighbors. \footnote{We here implemented interactions with $\nu = 2$ nearest neighbors in $1d$ and with $\nu = 4$ nearest neighbors in $2d$.}. 
In contrast to common dynamical formulations of the spatial prisoner's dilemma~\cite{NowakSpatial,Nowak:1994,Nakamuru:1997,Roca:2009a}, we drop the assumption of a coupled payoff-collection and selection step and thereby take into account that both processes in general occur on vastly different time-scales. Taken together, our model considers three different reaction steps which can occur at distinct rates. We chose our time unit $\tau = 1$ as the average lifetime of an individual when the selection is neutral (i.e.~when the fitness of all individuals is $1$), see below \footnote{As will be pointed out later, selection can only happen in the model if the other species is present in the neighborhood. Precisely, $\tau=1$ corresponds to the average lifetime if an individual is completely surrounded by neighbors with the opposing strategy and if the selection is neutral}. The rates are then defined as the average number of steps occurring per unit time $\tau$.

\emph{Payoff collection} occurs at a per capita rate $r$.  An individual engages in a pairwise interaction with one of its randomly chosen nearest neighbors. When the chosen neighbors are distinct, the cooperator $C$ collects a payoff $-c$ due to the costs of cooperation while the free-rider $F$ receives the benefit $b>c>0$. When the neighbors are both cooperators or both free-riders they receive a payoff $b-c$ and $0$, respectively. The payoff collection process of each individual continues during the whole course of its life, and the \emph{collected fitness}  $f_j$ of an individual $j$ at a certain time is given by the collected payoff the individual has received up to that time,
\begin{equation}
f_j = 1 + \tfrac{1}{n}\sum_{l=1}^n p_j^{(l)}.
\label{fitnessaccumulation}
\end{equation}
Here, $n$ denotes the number of interactions individual $j$ encountered. $p_j^{(l)}$ denotes the payoff for individual $j$ received during the $l$-th interaction with another individual. In addition, a background fitness of $1$ describes impacts on the fitness which are the same for each individual. We also implemented an alternative model, in which only the last $n_\text{max}$ interaction steps contribute to the fitness. This is a possible implementation for a biological situation in which interactions from the distant past do not contribute to the present fitness; $n_\text{max}$ can be seen as a memory range. A short memory range can significantly change the outcome, but if $n_\text{max} \gg 1$, the results are equivalent to the model without memory limitation. The impact of this alternative model of fitness collection will be further discussed below.

For the rest of the paper, we take a fixed collection rate $r\equiv 5$, i.e. the collection rate is chosen significantly larger than the reproduction timescale. Actually, for many biological situations, $r \gg 1$ is expected \cite{Diggle}. However, our qualitative findings are robust under the choice of $r$.

\emph{Reproduction and selection} is given by the replacement of one individual $j$ by a new individual belonging to the other type. Importantly, the rate for such a replacement event to occur depends on the type of the individual $j$, and the fitness of its nearest neighbors. For specificity, we take the corresponding transition rate for an individual $j$ at a site $x_j$ as the fitness of its neighbors $i \in\lbrace 1,...,\nu\rbrace$ with the opposite strategy times the probability ${1}/{\nu}$ of an interaction with them:
\begin{subequations}
\label{transitionrates}
\begin{eqnarray}
a_{C \rightarrow F}(j) = s_j \frac{1}{\nu}\sum_{i=1}^\nu f_i \cdot(1 - s_i),\\
a_{F \rightarrow C}(j) = (1 - s_j) \frac{1}{\nu}\sum_{i=1}^\nu f_i\cdot s_i.
\end{eqnarray}
\end{subequations}
Here, $s_l = 0$ individual $l$ is a free-rider and $s_l = 1$ if individual $l$ is a cooperator. If the neighbor at site $x_i$ belongs to the same type as the individual at site $x_j$, this neighbor does not contribute to the replacement rate $a_{\square \rightarrow \square}(j)$. Conversely, if it belongs to the other type, it increases the replacement rate of $j$ by its fitness $f_i$. If a selection event has taken place, we assume that the new individual starts with zero fitness \footnote{Alternatively, a new individual could start with the neutral fitness of $1$. However, the model was robust under this change}.

\emph{Mobility} is taken to be independent of fitness collection and selection dynamics. An individual interchanges its position with one of its randomly chosen nearest neighbors with a per capita hopping rate $M$. If $\ell$ is the lattice spacing, this corresponds to a macroscopic diffusion constant of $D=\ell^2 M$ in one dimension, and $D = \tfrac{1}{2} \ell^2 M$ in two dimensions. While payoff-collection and mobility events per individual occur with constant rates $r$ and $M$, respectively, the rate for a selection event, $a_{C\to F}(j) + a_{F\to C}(j)$, depends both on the life history of the interacting individuals and the composition of the local neighborhood. The time evolution of the system is determined by a Markov process ensuing from the rates for the various processes introduced above. To investigate the dynamics we performed extensive stochastic simulations of the underlying time-continuous stochastic process employing the Gillespie algorithm~\cite{Gillespie}.

\section{Simulation Dynamics and Qualitative Behaviour}
In the following, we first consider the evolutionary dynamics for the limits of vanishing and large mobility before considering intermediate mobilities and the ensuing absorbing state phase transition.

In the limit of large mobility, $M \rightarrow \infty$, the population is well-mixed and spatial degrees of freedom can be neglected. In particular, an individual interacts with all other individuals with equal probability and thus samples payoff values from interactions with the whole population. Hence, for a given overall fraction $x$ of cooperators in the population, the expected fitness value $f_j$ of an individual $j$ is site-independent, and for cooperators resp. free-riders given by 
\begin{subequations}
\label{eq:expected_fitness}
\begin{eqnarray}
f_C = 1 + x(b-c)  + (1-x)(-c), \\
f_F = 1 + x b.
\end{eqnarray}
\end{subequations}
Replacement of individuals occurs according to the expected replacement rates. With Eqs.~\eqref{transitionrates}, they are given by $\langle a_{C \rightarrow F} \rangle = x(1-x) f_F $ and $\langle a_{F \rightarrow C} \rangle = x(1-x) f_C $. The expected change of the global fraction of cooperators is thus given by
\begin{equation}
\label{eq:mean}
\partial_t x =\langle a_{F \rightarrow C} \rangle -\langle a_{C \rightarrow F} \rangle = -c x (1-x).
\end{equation}
This is the common replicator dynamics of the prisoner's dilemma in a well-mixed population~\cite{Maynard}; due to the costs $c$ of cooperation only free-riders survive in the long run. For a more complete description of the well-mixed system, fluctuations due to the finite size of the system (finite $N$) and an effectively limited payoff collection step (finite $n$) can be taken into account~\cite{Nowak:2004,Traulsen:2005,Cremer:2009,Galla:2009,CremerPRE:2011}.

\begin{figure}
\centering \includegraphics[width= \columnwidth]{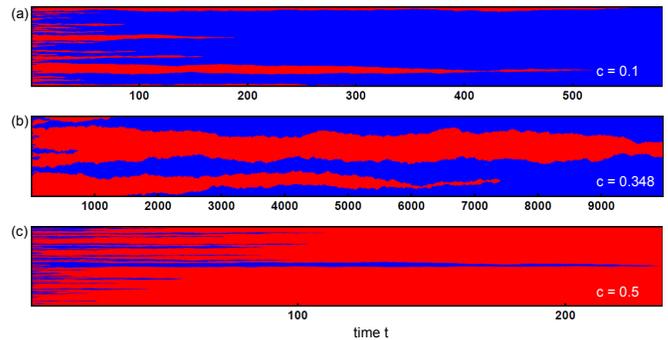}
\caption{
\label{fig:timeevolution0} 
(Color online) Lattice occupation (vertical axis) versus time (horizontal axis) of the one-dimensional dynamics for vanishing mobility, $M=0$. The lattice size was $N=500$. Starting with a random distribution of $50 \%$ cooperators [dark grey (blue)] and $50 \%$ free-riders [light grey (red)] on the lattice, clusters are formed. Preferentially, the clusters of one type grow, such that only cooperators or only free-riders remain in the end. (a) the cooperation costs $c$ are $0.1$. For these low costs cooperators take over the population. (b) $c=0.347$, the cooperation costs are very close to the critical value $c_0$. The cluster extension and the cluster lifetime both show a power-law divergence as $c \rightarrow c_0$ for increasing $N$. (c) $c=0.5$, for high costs cooperators die out and only free-riders survive.}
\end{figure}
\begin{figure}
\centering \includegraphics[width=\columnwidth]{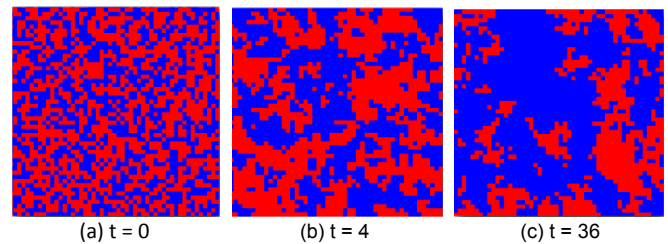}
\caption{
\label{fig:clusters}
(Color online) Snapshots of the two-dimensional lattice and cluster formation for vanishing mobility, $M=0$ at different times $t$. Starting with a random distribution of half cooperators [dark grey (blue)] and half free-riders [light grey (red)] on the lattice (a), clusters are formed (b). Preferentially, the clusters of one type grow (c), such that only cooperators or only free-riders remain in the end.  This simulation was performed on a $100 \times 100$ lattice, the cooperation costs were $c = 0.1$.}
\end{figure}
\begin{figure}[htb]
\centering \includegraphics[width= 0.9\columnwidth]{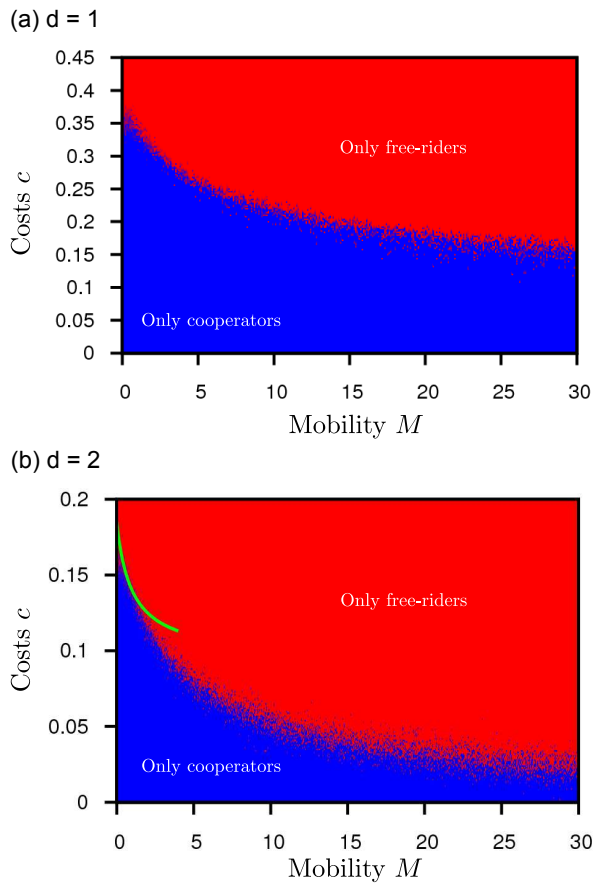}
\caption{
(Color online) Stationary fraction $x$ of cooperators as a function of the costs of cooperation $c$ and the mobility $M$ for (a) the one-dimensional and (b) the two-dimensional model. There is a phase transition between stationary states where only cooperation [dark grey (blue)] or only non-cooperation [light grey (red)] prevails. $N=500$ in (a), $N=100$ in (b). 
\label{fig:phasediag}}
\end{figure}

Here we consider the fully stochastic dynamics of the spatially extended system. Fig.~\ref{fig:timeevolution0} shows typical time evolution scenarios of the one-dimensional game for the limit of vanishing mobility, $M=0$, and Fig.~\ref{fig:clusters} depicts different snapshots of the two-dimensional dynamics in the same limit. When starting with a random initial configuration of the lattice with an average cooperator fraction of $x_0=0.5$, first compact clusters of cooperators and free-riders are formed after few generations. As implied by the definition of the selection rates, Eq.~\eqref{transitionrates}, there are no transitions between different types within clusters but only at the domain boundaries between cooperator and free-rider clusters. There, the dynamics is determined by two antagonistic effects. On the one hand, cooperators at the boundary benefit from the neighboring cooperators within their cluster, and from their previous interactions. On the other hand, a free-riding individual still has a fitness advantage by saving the costs for providing cooperation. From our numerical simulations we observe that for a benefit $b = 1$, the costs of cooperation have to be lower than a critical value, $c_0 = 0.3471 \pm 0.0008$ in one dimension and $c_0 = 0.163 \pm 0.0005$ in two dimensions, for cooperators to persist, see Fig.~\ref{fig:phasediag}~(a) and (b). Throughout the rest of this manuscript we take the benefit $b=1$ as fixed. Other values for $b$ do not lead to qualitative changes of the model.

For the alternative implementation of the payoff collection, where only the last $n_\text{max}$ payoff collection steps contribute to the fitness, the qualitative dynamics remained unchanged as long as $n_\text{max}$ was greater than $1$. But importantly, cooperators always died out in one dimension if only the last payoff collection step was remembered ($n_\text{max}=1$). Therefore, in a one-dimensional system the memory of past interactions is an essential mechanism for the support of cooperation, and spatial structure alone cannot promote cooperation: in one dimension, both a cooperator and a free-rider at a domain wall always interact with exactly one cooperator and one free-rider. If just the payoff from such a configuration is remembered, the free-rider at the domain wall will on the average always be fitter than the cooperator (if the cooperation costs $c$ are greater than zero). However, if past interactions are remembered, a cooperator that has been inside of its cluster for a long time has acquired a high average payoff from these past interactions. Such a cooperator will not suffer too much from few interactions with a free-rider when it comes to a domain wall. In turn, a free-rider that has been within its cluster very long has also a very low fitness. Memory effects can therefore strengthen cooperation and are crucial for the cooperator survival in one dimension. In two dimensions, on the other hand, spatial structure alone can maintain cooperation, even without memory effects (i.e.~even if $n_\text{max}=1$). A cooperator at a domain wall can get a benefit from up to three cooperating neighbors from its cluster if the domain wall is linear. An advantageous domain wall can make mutual cooperation beneficial and protect cooperators if the costs $c$ are not too high.

However, cooperation is unstable under the influence of mobility, both in one and in two dimensions. The interchange in position of neighboring individuals disturbs the formation of sharply separated clusters. With increasing mobility $M$, cluster boundaries become more and more blurred. As a consequence, cooperators are surrounded by fewer and fewer peers and thereby lose their advantage of interacting in a highly cooperative neighborhood. Cooperativity therefore strongly decays when $M$ is increased. Remarkably, we find that there is a critical phase transition separating phases where only cooperators or only free-riders are expected to persist in the long run. While for sufficiently high mobility and costs only free-riders remain, cooperation is the only surviving strategy when mobility and costs fall below critical values. Thus in either case the stationary state of the dynamics is an absorbing state. The phase diagrams for the one- and for the two-dimensional model are shown in Fig.~\ref{fig:phasediag}(a) and (b), respectively. Other than in previous Prisoner's Dilemma models~\cite{Szabo:1998, Szab'o:2002, Szabo:2005}, no fluctuating active phase with coexistence of cooperators and free-riders was observed. The decay of the critical costs $c_M$ with $M$ is found to be very strong for unbiased mobility(cf.~Fig.~\ref{fig:phasediag}(a) and (b)), threatening cooperation even if each individual performs only a couple diffusion steps per lifetime (i.e.~if $M$ is only $O(1)$ and comparable in size with the reproduction rate).

\section{Cluster Approximation for Critical Costs}

\begin{figure}[htb]
\centering \includegraphics[width = \columnwidth]{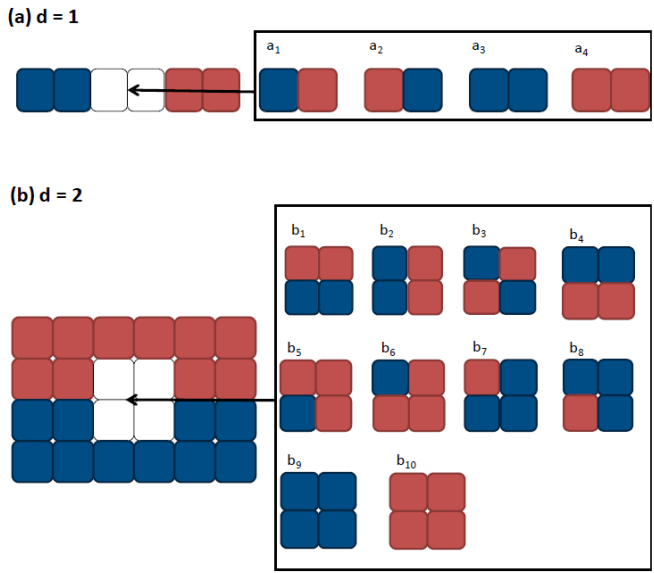}
\caption{
\label{fig:frontier} 
(Color online) Illustration of a simplified domain wall picture for the one-dimensional (a), and two-dimensional (b) lattice. The domain wall dynamics is constrained to the white area where transitions between all possible arrangements of cooperators [dark grey (blue)] and free-riders [light grey (red)] are possible, as indicated in the graphs.The configuration of cooperators and free-riders outside of the white area is considered to remain static. The one-dimensional domain wall has moved forward or backward if the dynamic sites in the white area are in the all-cooperator state $a_3$ or the all-free-rider state $a_4$, respectively. Similarly, the section considered for the two-dimensional domain wall has moved forward or backward if the dynamic sites are in states $b_9$ or $b_{10}$, respectively.}
\end{figure}

The population dynamics is mainly governed by processes occurring at the boundaries between domains dominated by cooperators and free-riders, respectively.  In this section, we give a simplified analysis for the dynamics of theses domain walls and thereby approximately calculate the critical costs $c_0$ below which cooperators can survive. The schematics of the simplified domain wall picture is shown in Fig.~\ref{fig:frontier}(a) and (b) for the  one- and two-dimensional case, respectively. For the one-dimensional case we consider an isolated domain wall, and reduce the dynamics to the two boundary sites between a cooperator and a free-rider domain, and consider the remainder of both domains to be static. The possible configurations of the domain wall boundary are illustrated by Fig.~\ref{fig:frontier}~$a_1$--$a_4$. Among others, this scheme neglects events where domain walls are created or annihilated.
Similarly, for the two-dimensional lattice, we constrain the dynamics to the immediate vicinity of the domain wall and consider a $2\times2$ section at the domain wall boundary whose possible configurations are shown in Figs.~\ref{fig:frontier}~$b_1$--$b_{10}$. The remainder of the domain wall is considered as a static horizontal front. Again, these assumptions leave out many other possible processes changing the domain wall boundaries.  Nevertheless, such a strongly simplified model seems to retain the most essential features of the domain wall dynamics, since the estimates for the critical costs $c_0$ are in good agreement with the simulation results (see below).

We start our analysis with the one-dimensional model. The dynamics of the domain wall is then governed by transitions between the four states $a_1$, ..., $a_4$, which occur at rates determined by Eqs.~(2a) and (2b) and the history of each individual. We resort to a mean-field picture and aim to derive a rate equation, $\dot{\vec{v}} = T \cdot \vec{v}$, which determines the time evolution of the probabilities $\vec{v} := (a_1(t), ..., a_4(t))$ to find one of these four configurations. Let us first assume that each individual collects the payoff from all of its neighbors, such that the fitness of each individual can be approximated by the mean payoff received from its neighborhood. The transition rates between different configurations can then be calculated from Eq.~\eqref{transitionrates}. For example, the transition from $a_2$ to $a_3$ 
occurs if the free-rider in configuration $a_2$ turns into a cooperator. According to Eq.~\eqref{transitionrates}, the rate for this to happen is the fitness of the cooperating left and right neighbors, $f_\text{left/right}$, times the probability $\tfrac{1}{2}$ that a particular neighbor is chosen:
\begin{eqnarray}
T_{32} = \tfrac{1}{2} \left( f_\text{left} + f_\text{right} \right) 
            = \tfrac{5}{4} - c .
\end{eqnarray}
Under the assumption, that these neighbors have previously collected equal amounts of fitness from both of its neighbors ($1$ is the neutral fitness, $1-c$ is the payoff from a cooperating neighbor, $-c$ is the payoff from a free-riding neighbor), the fitness for the right and left neighbor can be calculated. Since the left neighbor is surrounded by a cooperator and a free-rider its collected fitness is $f_\text{left} = 1 + \tfrac{1}{2} \left[ (1-c) + (-c) \right]$. Similarly, since the right neighbor is surrounded by two free-riders it gets a payoff of $-c$ from both of them such that $f_\text{right}=1 +  \tfrac{1}{2} \left[ (-c) + (-c) \right]$. Analogous calculations for the remaining transitions rates lead to the following transition matrix $T$:
\begin{equation}
T = \left(
\begin{array}{cccc}
-\tfrac{3}{2} +\tfrac{c}{2}&0&\tfrac{3}{4}&\tfrac{3}{4}-\tfrac{c}{2}\\
0&-3+c&0&0\\
\tfrac{3}{4}-\tfrac{c}{2}&\tfrac{5}{4} - c&-\tfrac{3}{4}&0\\
\tfrac{3}{4}&\tfrac{7}{4}&0&-\tfrac{3}{4}+\tfrac{c}{2}
\end{array} \right).
\end{equation}
The stationary probability vector is given by $T \cdot \vec{v}^* = 0$. Whether the domain wall mainly moves forward or backward is determined by the difference between the stationary values of the all-cooperator state $a_3$ and the all-free-rider state $a_4$. The sign determines whether cooperators or free-riders finally take over the whole population. Calculating the null-vector $\vec{v}^*$ we find that $a_3-a_4 = \tfrac{1}{9}(4 c^2 - 12 c)$ which is strictly negative unless $c=0$. Therefore, if we can assume that the fitness of each individual is given by the mean payoff, the critical costs for cooperation is $c_0 = 0$. This result reemphasizes that spatial structure alone cannot promote cooperation in one dimension. However, as we know from our numerical analysis in the previous section, memory effects -- not taken into account so far -- can actually promote cooperation.

Within the framework of the simple model depicted in Fig.~\ref{fig:frontier} infinite memory can be incorporated by assigning to the individuals in the static clusters a fixed fitness of $2-c $ and $1$ for cooperators and free-riders, respectively. The underlying assumption is that these static players have been inside of their respective cluster for a long time and have not interacted much with the opposing strategy. For example, consider a static cooperator that has been inside of its cluster during the last $n-1$ ($n\gg 1$) interaction steps, and finally ends up at the domain wall, where it interacts with a free-rider once. The fitness of such a cooperator would then according to Eq.~\eqref{fitnessaccumulation} be given by $f_C = 1 + \tfrac{1}{n} [(n-1)(1-c) - c]$. If the number of previous interactions within a cluster is assumed to be very large for individuals in static clusters, their fitness can therefore be approximated $2-c$ and $1$ for cooperators and free-riders, respectively. With this assumption for static players, the transition rate from $a_2$ to $a_3$ now reads $\tfrac{1}{2}(2-c) + \tfrac{1}{2}(1 + \tfrac{-c}{2} + \tfrac{-c}{2}) = \tfrac{3}{2} - c$. This is similar to the above calculation but the static cooperator left of the free-rider in $a_2$ is now assumed to have a fitness of $2-c$. With this modification, the transition matrix becomes
\begin{equation}
T = \left(
\begin{array}{cccc}
-\tfrac{3}{2} +\tfrac{c}{2}&0&\tfrac{1}{2}&1-c\\
0&-3+c&0&0\\
\tfrac{3}{4}-\tfrac{c}{2}&\tfrac{3}{2} - c&-\tfrac{1}{2}&0\\
\tfrac{3}{4}&\tfrac{3}{2}&0&-1+c
\end{array} \right) .
\end{equation}
Now we find that in the stationary state $a_3-a_4 = 1 -  \tfrac{10}{3}c + \tfrac{4}{3} c^2$, which implies  a critical cost of $c_0=\tfrac{5 - \sqrt{13}}{4} \approx 0.3486$. This value is in good agreement both with the phase diagram Fig.~3(a) and with the numerical value for the critical costs determined by finite-size scaling (see table~1).

In two dimensions, spatial structure alone can provide a mechanism for the survival of cooperation and therefore memory effects do not need to be taken into account. The calculation of the transition rates between the ten different configurations illustrated in Fig.~\ref{fig:frontier}(b) is analogous to the one-dimensional case. For example, a transition between $b_1$ and $b_5$ occurs if either one of the two cooperators in configuration $b_1$ turns into a free-rider. In each of these two cases there is a probability of $\tfrac{1}{4}$ that for the given cooperator a free-rider is chosen as an interaction partner. The fitness of which is $1 + \tfrac{1}{4}$ since it has one cooperator but three free-riders as its neighbors. Taken together, the transition rate reads $ T_{51}= 2 \, \tfrac{1}{4}(1 + \tfrac{1}{4}) = \tfrac{5}{8}$. Transitions between the various configurations $b_i$ may also occur due to hopping events. These were disregarded in our discussion of the one-dimensional case. Here, we take those events into account in the limit of small mobility rates; for large mobilities clusters become more and more blurred and the cluster approximation is not applicable. As an example, consider the transition from $b_7$ to $b_8$, which is a purely diffusive transition. The transition rate is $\tfrac{2}{4} M$, where $2 M$ is the rate at which either the free-rider in $b_7$ or its lower cooperating neighbor are chosen for a mobility step, and $\tfrac{1}{4}$ is the probability that the other neighbor (the free-rider, if the cooperator was previously chosen, and vice versa) is picked for swapping places
\begin{widetext}
\arraycolsep=-0.2pt
\medmuskip = 1mu
\noindent\hspace{-20mm}
\begin{equation}
T= \left(
\scalemath{0.73}{
\begin{array}{cccccccccc}
-\tfrac{3}{2} + \tfrac{c}{2} - M&0&\tfrac{M}{2}&0&\tfrac{19}{16} - \tfrac{3 c}{4}&0&\tfrac{17}{16}&0&0&0\\
0&-3 + c - M&\tfrac{M}{2}&0&\tfrac{3}{8} - \tfrac{c}{4}&\tfrac{17}{16} - \tfrac{3 c}{4}&\tfrac{3}{8}&\tfrac{19}{16}&0&0\\
M&M&- \tfrac{41}{8} + \tfrac{3 c}{2} - \tfrac{3 M}{2}&M&0&\tfrac{13}{16} - \tfrac{c}{2}&0&\tfrac{13}{16}&0&0\\
0&0&\tfrac{M}{2}&- \tfrac{9}{2} + \tfrac{3 c}{2} - M&0&\tfrac{1}{4} - \tfrac{c}{4}&0&\tfrac{1}{2}&0&0\\
\tfrac{5}{8}&1&\tfrac{25}{16}&0&- \tfrac{37}{16} + c - \tfrac{M}{2}&\tfrac{M}{2}&0&0&0&\tfrac{13}{8} - c\\
0&\tfrac{7}{16}&\tfrac{7}{8}&\tfrac{9}{4}&\tfrac{M}{2}&-\tfrac{57}{16} + \tfrac{3 c}{2} - \tfrac{M}{2}&0&0&0&0\\
\tfrac{7}{8} - \tfrac{c}{2}&\tfrac{5}{4} - \tfrac{3 c}{4}&\tfrac{23}{16} - c&0&0&0&- \tfrac{35}{16} + \tfrac{c}{2} - \tfrac{M}{2}&\tfrac{M}{2}&\tfrac{11}{8}&0\\
0&\tfrac{5}{16} - \tfrac{c}{4}&\tfrac{5}{4} - \tfrac{c}{2}&\tfrac{9}{4} - \tfrac{3 c}{2}&0&0&\tfrac{M}{2}&- \tfrac{65}{16} + c - \tfrac{M}{2}&0&0\\
0&0&0&0&0&0&\tfrac{3}{4} - \tfrac{c}{2}&\tfrac{25}{16} - c&- \tfrac{11}{8}&0\\
0&0&0&0&\tfrac{3}{4}&\tfrac{23}{16}&0&0&0&-\tfrac{13}{8}+c
\end{array}
}
\right) \nonumber
\end{equation}
\normalsize

\end{widetext}
The condition that no movement of the front occurs in the stationary state, i.e.~that $b_9 - b_{10} = 0$, leads to an approximate numerical solution for the critical costs $c_M$, which is shown as a green line in Fig.~2(b).

\section{Critical Behavior at the Absorbing State Phase Transition}

In the following, we investigate the critical behavior at the absorbing state phase transition. To this end, we analyze the formation, dynamics and annihilation of clusters. We recorded the life-time $T_\alpha$ of each cluster $\alpha$, as well as the maximum linear extension (arbitrary direction),  $L_\alpha$, during its life-time. These measured lengths and lifetimes are expected to scale the same way as the characteristic lengths and times of the system \cite{Hinrichsen:2000}. We further logged the cumulative mass $m_\alpha$ of each cluster during its life time, i.e.~the number of individuals belonging to the cluster at focus, integrated over its whole life-time. Initial conditions where such that each site was randomly occupied by either a cooperator or a free-rider with the same probability $0.5$. All measurements were performed for clusters doomed to extinction: For the phase where only cooperators prevail, we regarded only free-riding clusters and vice versa.

\begin{figure*}[htb]
\centering
\centering \includegraphics[width = \textwidth]{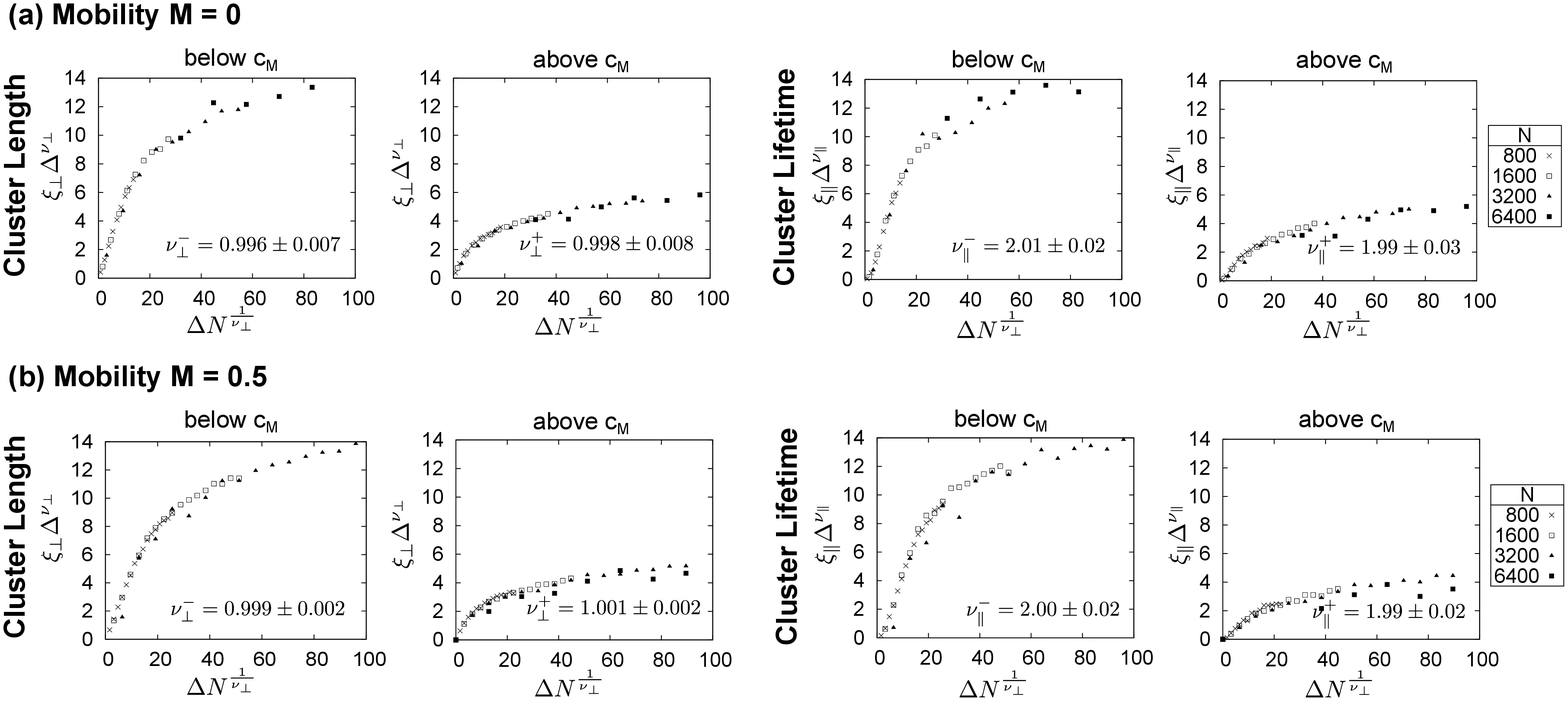}
\caption{
\label{fig:finitesize1d}
One-dimensional model: Finite size scaling for varied costs $c$ at (a) $M = 0$, (b) $M = 0.5$. $\Delta=|c-c_M|$. With the determined exponents $\nu^\pm_\parallel$ and $\nu^\pm_\perp$, and the critical costs [$c_M=0.347 \pm 0.0008$ in case (a) and  $c_M=0.3620 \pm 0.0004$ in case (b)], the data for $\xi_{\parallel}$ and $\xi_{\perp}$ at different system sizes collapsed onto the master curves $\mathcal{F}^\pm$ and $\mathcal{G}^\pm$, see main text. }
\end{figure*}

\begin{figure*}[htb]
\centering \includegraphics[width = \textwidth]{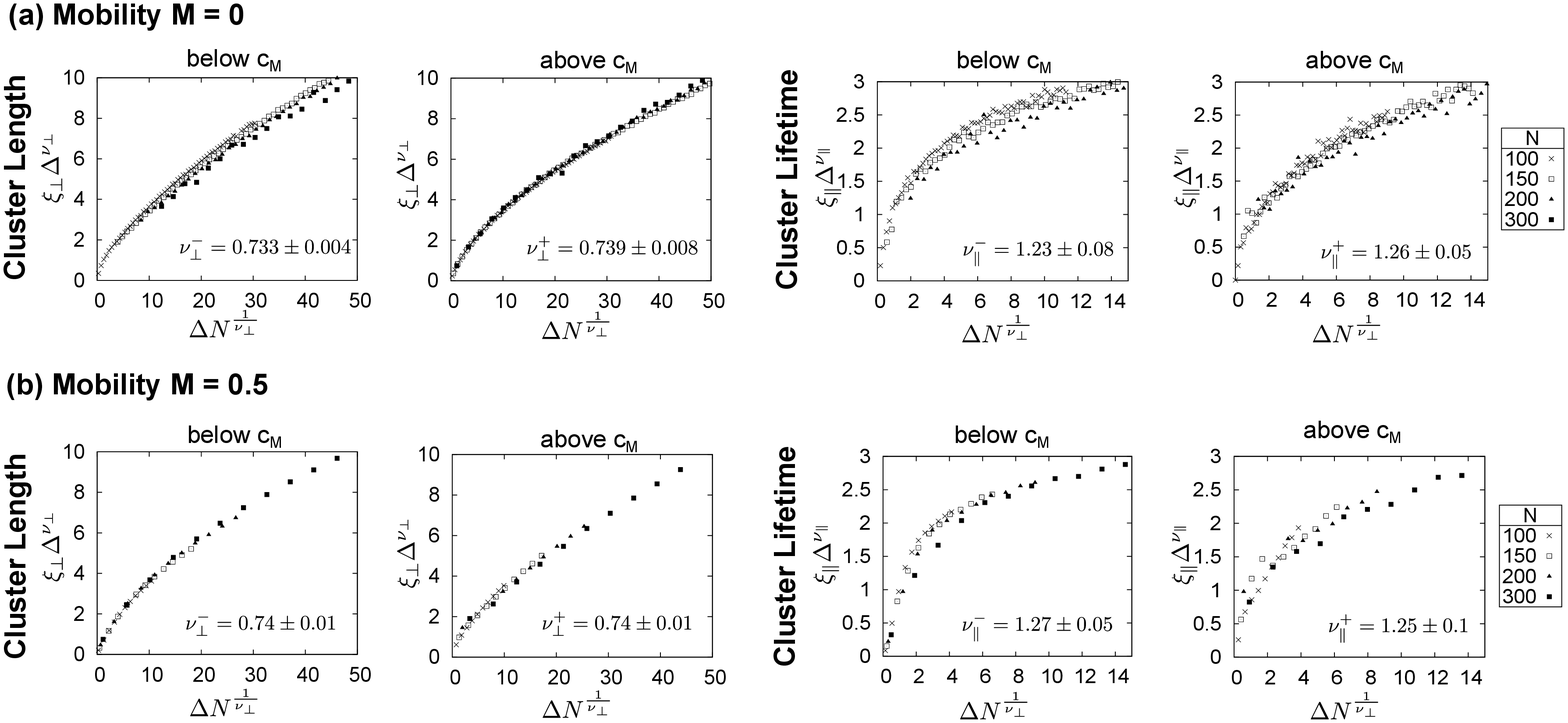}
\caption{
\label{fig:finitesize2d}
Two-dimensional model: Finite size scaling for varied costs $c$ at (a) $M = 0$, (b) $M = 0.5$. $\Delta=|c-c_M|$. With the determined exponents $\nu^\pm_\parallel$ and $\nu^\pm_\perp$, and the critical costs [$c_M=0.163 \pm 0.0005$ in case (a) and  $c_M=0.1605 \pm 0.0007$ in case (b)], the data for $\xi_{\parallel}$ and $\xi_{\perp}$ at different system sizes collapsed onto the master curves $\mathcal{F}^\pm$ and $\mathcal{G}^\pm$, see main text.}
\end{figure*}

In order to obtain the typical temporal extension, $\xi_{\parallel}$, and the typical linear extension, $\xi_{\perp}$, we weighted the measured maximal extensions of all clusters $\alpha$ with their respective cluster masses~\cite{Stauffer:1992},
\begin{eqnarray}
\xi_{\parallel} \sim \frac{\sum_{\alpha} m_\alpha T_\alpha}{\sum_{\alpha} m_\alpha} \text{, and~ }
\xi_{\perp} \sim \frac{\sum_{\alpha} m_\alpha L_\alpha}{\sum_{\alpha} m_\alpha}.
\end{eqnarray}
Close to the phase transition, the typical time and length scales are expected to behave like power laws  with certain exponents $\nu_{\parallel}$ and $\nu_{\perp}$~\cite{Hinrichsen:2000}, respectively. However, due to the finite system size, these typical scales do not diverge at the critical point but, close to the critical point, are cut-off by finite size effects. Close to the critical point and for large enough system sizes one expects the following finite size scaling behavior for  $\xi_{\parallel}$ and $\xi_{\perp}$~\cite{Hinrichsen:2006}:
\begin{subequations}
\begin{eqnarray}
\xi_{\parallel} &\sim&
 \Delta ^{-\nu_\parallel} \mathcal{F}^\pm(\Delta N^\frac{1}{\nu_\perp}),\\
\xi_{\perp} &\sim&
\Delta ^{-\nu_\perp} \mathcal{G}^\pm(\Delta N^\frac{1}{\nu_\perp}).
\end{eqnarray}
\label{eq:finitesizescaling}
\end{subequations}
Here, $\Delta:=|c-c_M|$ denotes the distance from the mobility-dependent critical costs $c_M$, and $\mathcal{F}^\pm$ and $\mathcal{G}^\pm$ signify the scaling functions below $(-)$ and above $(+)$ the critical point.
 
As discussed before, the costs $c$ and the mobility $M$ control the phase transition, cf.~Fig.~\ref{fig:phasediag}. We analyzed the phase transition for fixed mobilities $M$ under variation of the costs $c$. We measured $\xi_{\parallel}$ and $\xi_{\perp}$ for different lattice sizes and used the results to estimate the critical exponents $\nu_{\parallel}$ and $\nu_{\perp}$. We achieved this by using finite size scaling, Eqs.~\eqref{eq:finitesizescaling}, \emph{i.e.} by adjusting the exponents $\nu_\parallel$, $\nu_\perp$ and the critical point $c_M$ to optimize data collapse. In one dimension we performed simulations for linear lattice sizes $N = \lbrace 800, 1600, 3200, 6400 \rbrace$, and in two dimensions for $N = \lbrace 100, 150, 200, 300 \rbrace$: the ensuing scaling functions are shown in  Fig.~\ref{fig:finitesize1d}  and Fig.~\ref{fig:finitesize2d} for the one-dimensional and two-dimensional model, respectively. We then determined the critical exponents $\beta$ and $\beta '$ by fitting the cooperator fraction $x(t)$ to the relation 
\begin{equation}
x(t) \sim t^{-\beta/\nu_\parallel}
\end{equation} 
and the survival probability $P(t)$ to the relation 
\begin{equation}
P(t) \sim t^{-\beta '/\nu_\parallel}, 
\end{equation}
which are valid at the critical point $c_M$~\cite{Hinrichsen:2006}.

We roughly estimated the error of the exponents by varying the exponents in the fitting procedure, considering the scaling functions for small and large arguments: For the optimal fitting, all data points lie within certain intervals around the (approximately) linear scaling functions, whereas for the estimated errors only $60\%$ of the points lie within at least one of those intervals. Our numerical results are summarized in table~1.
\begin{table}[h]
\centering \begin{tabular}{|c|c|c|c|c|}
\hline  & \multicolumn{2}{|c|}{$d=1$} &  \multicolumn{2}{|c|}{$d=2$}\\
\hline  &$M=0$& $M=0.5$ & $M=0$ & $M=0.5$ \\ 
\hline $c_M$ & $0.347 \pm 0.001$ & $0.362 \pm 0.001$ & $0.163 \pm 0.001$ & $0.161 \pm 0.001$ \\ 
\hline $\nu^-_{\parallel}$ & $2.01 \pm 0.02$ & $2.00 \pm 0.02$ & $1.23 \pm 0.08$ & $1.27 \pm 0.05$ \\ 
\hline $\nu^+_{\parallel}$ & $1.99 \pm 0.03$ & $1.99 \pm 0.02$ & $1.26 \pm 0.05$ & $1.25 \pm 0.1$ \\ 
\hline $\nu^-_{\perp}$ & $0.996 \pm 0.007$ & $0.999 \pm 0.002$ & $0.733 \pm 0.004$ & $0.74 \pm 0.01$ \\ 
\hline $\nu^+_{\perp}$ & $0.998 \pm 0.008$ & $1.001 \pm 0.002$ & $0.739 \pm 0.008$ & $0.74 \pm 0.01$ \\ 
\hline $\beta$ & $0.02 \pm 0.02$ & $0.04 \pm 0.04$ & $0.56 \pm 0.03$ & $0.56 \pm 0.09$ \\ 
\hline $\beta '$ & $1.01 \pm 0.02$ & $0.92 \pm 0.2$ & $0.57 \pm 0.08$ & $0.6 \pm 0.1$ \\ 
\hline
\end{tabular} 
\label{tab:results}
\caption{Summary of the measured critical exponents in one and two dimensions, and for two different mobility rates, $M=0$ and $M=0.5$.}
\end{table}

The measured exponents indicate that the phase transition belongs to the voter model universality class in one dimension, where the scaling exponents corresponding to the population fraction, the survival probability, the cluster size and the temporal correlations are given by $\beta = 0$, $\beta' = 1$, $\nu_\parallel = 2$ and $\nu_\perp = 1$ in one dimension~\cite{Dickman:1995, Ben-Naim:1995, Odor:2004}. In fact, there are many similarities to the Voter Model, like two absorbing phases, compact clusters and the absence of an active phase. The two-dimensional system apparently belongs to the universality class of directed percolation (DP). For this class, $\beta = \beta ' = 0.584(4)$, $\nu_\parallel = 1.295(6)$ and $\nu_\perp = 0.734(4)$ \cite{Hinrichsen:2006}. Interestingly, in contrast to DP, the dynamics violates the condition of a unique absorbing state. Moreover, in contrast to other known DP class models~\cite{Janssen:1981, Grassberger:1982}, the model does not show a fluctuating active phase. The existence of an active phase seems not to be a necessary condition for the DP universality class.

\section{Summary and Outlook}

In summary, we have investigated the impact of mobility on the evolution of cooperation. Importantly, by introducing a separated fitness collection dynamics we accounted for the biological fact that fitness is the result of many underlying processes. Only by this fitness collection dynamics, the correct asymptotic limits for vanishing and large mobility, the spatial and the well-mixed variants of the prisoner's dilemma are obtained. While in two dimensions, spatial structure alone is sufficient to promote cooperation, the memory of past interactions is a necessary condition for cooperation in one dimension. In fact, memory effects can be seen as an additional mechanism favoring cooperation -- and they are expected to occur even in simple organisms, since their fitness depends on several interactions with their environment. Memory effects have already been found to promote cooperation in a deterministic game \cite{Qin:2008, AlonsoSanz:2009}. The present study confirms this finding for a stochastic setup and shows that they are of essential importance in one dimension.

For a certain intermediate mobility (depending on benefit and costs) there is a critical phase transition without stable coexistence, both in one and in two dimensions. Below critical mobilities and costs only cooperators remain while above only free-riders remain in the long run. This phase transition is robust against changes of dynamical details, like a limited payoff collection capability. 

More importantly, for cooperation to prevail in the spatial prisoner's dilemma, the time-scale of mobility must always be of the same order or lower as the selection time-scale. If one considers for example microbial populations, the spatial prisoner's dilemma can only explain cooperation if the reproduction time of the microbes under consideration is of the same order as the time they need to move to a neighboring bacterium. This condition is probably not fulfilled in most ecological situations. Thus, at least in its standard formulation, the spatial prisoner's dilemma might serve as a placative example to explain how spatial clustering can promote cooperation in principle. But it cannot serve as a substantive explanation for cooperative behavior in natural populations. In microbial populations, more complex clustering of microbes into different colonies, the coupling to growth dynamics, and the dynamical restructuring of the population on larger length scales are probably more important for the evolution and sustainment of cooperative strains \cite{Hamilton:1964, Fletcher:2006, Rainey:2003, LeGac:2009, Cremer:2012, Melbinger:2010, West:2007b,Hallatschek:2007,Hallatschek2008158,Kuhr:2011,Korolev2011,Lavrentovich2013}.

\begin{acknowledgments}
We thank Jan-Timm Kuhr, Anna Melbinger and Noreen Walker for discussions. 
Financial support of Deutsche Forschungsgemeinschaft through the German Excellence Initiative via the program ``Nanosystems Initiative Munich'' (NIM) and through the SFB TR12 ``Symmetries and Universalities in Mesoscopic Systems'' is gratefully acknowledged.
\end{acknowledgments}


\begin{thebibliography}{60}
\expandafter\ifx\csname natexlab\endcsname\relax\def\natexlab#1{#1}\fi
\expandafter\ifx\csname bibnamefont\endcsname\relax
  \def\bibnamefont#1{#1}\fi
\expandafter\ifx\csname bibfnamefont\endcsname\relax
  \def\bibfnamefont#1{#1}\fi
\expandafter\ifx\csname citenamefont\endcsname\relax
  \def\citenamefont#1{#1}\fi
\expandafter\ifx\csname url\endcsname\relax
  \def\url#1{\texttt{#1}}\fi
\expandafter\ifx\csname urlprefix\endcsname\relax\def\urlprefix{URL }\fi
\providecommand{\bibinfo}[2]{#2}
\providecommand{\eprint}[2][]{\url{#2}}

\bibitem[{\citenamefont{Axelrod and Hamilton}(1981)}]{axelrod-1981-211}
\bibinfo{author}{\bibfnamefont{R.}~\bibnamefont{Axelrod}} \bibnamefont{and}
  \bibinfo{author}{\bibfnamefont{W.}~\bibnamefont{Hamilton}},
  \bibinfo{journal}{Science} \textbf{\bibinfo{volume}{211}},
  \bibinfo{pages}{1390} (\bibinfo{year}{1981}).

\bibitem[{\citenamefont{{Maynard Smith}}(1982)}]{Maynard}
\bibinfo{author}{\bibfnamefont{J.}~\bibnamefont{{Maynard Smith}}},
  \emph{\bibinfo{title}{Evolution and the Theory of Games}}
  (\bibinfo{publisher}{Cambridge University Press},
  \bibinfo{address}{Cambridge}, \bibinfo{year}{1982}).

\bibitem[{\citenamefont{Hamilton}(1964)}]{Hamilton:1964}
\bibinfo{author}{\bibfnamefont{W.~D.} \bibnamefont{Hamilton}},
  \bibinfo{journal}{J.Theor. Biol.} \textbf{\bibinfo{volume}{7}},
  \bibinfo{pages}{1} (\bibinfo{year}{1964}).

\bibitem[{\citenamefont{Nowak and May}(1992)}]{NowakSpatial}
\bibinfo{author}{\bibfnamefont{M.~A.} \bibnamefont{Nowak}} \bibnamefont{and}
  \bibinfo{author}{\bibfnamefont{R.~M.} \bibnamefont{May}},
  \bibinfo{journal}{Nature} \textbf{\bibinfo{volume}{359}},
  \bibinfo{pages}{826} (\bibinfo{year}{1992}).

\bibitem[{\citenamefont{Roca et~al.}(2009)\citenamefont{Roca, Cuesta, and
  Sanchez}}]{Roca:2009a}
\bibinfo{author}{\bibfnamefont{C.~P.} \bibnamefont{Roca}},
  \bibinfo{author}{\bibfnamefont{J.~A.} \bibnamefont{Cuesta}},
  \bibnamefont{and} \bibinfo{author}{\bibfnamefont{A.}~\bibnamefont{Sanchez}},
  \bibinfo{journal}{Phys. Life. Rev.} \textbf{\bibinfo{volume}{6}},
  \bibinfo{pages}{208} (\bibinfo{year}{2009}).

\bibitem[{\citenamefont{Durrett and Levin}(1994)}]{Durrett:1994}
\bibinfo{author}{\bibfnamefont{R.}~\bibnamefont{Durrett}} \bibnamefont{and}
  \bibinfo{author}{\bibfnamefont{S.}~\bibnamefont{Levin}},
  \bibinfo{journal}{Theoretical Population Biology}
  \textbf{\bibinfo{volume}{46}}, \bibinfo{pages}{363 } (\bibinfo{year}{1994}),
  ISSN \bibinfo{issn}{0040-5809}.

\bibitem[{\citenamefont{Nowak et~al.}(1994)\citenamefont{Nowak, Bonhoeffer, and
  May}}]{Nowak:1994}
\bibinfo{author}{\bibfnamefont{M.~A.} \bibnamefont{Nowak}},
  \bibinfo{author}{\bibfnamefont{S.}~\bibnamefont{Bonhoeffer}},
  \bibnamefont{and} \bibinfo{author}{\bibfnamefont{R.~M.} \bibnamefont{May}},
  \bibinfo{journal}{Proc. Natl. Acad. Sci. U. S. A.}
  \textbf{\bibinfo{volume}{91}}, \bibinfo{pages}{4877} (\bibinfo{year}{1994}).

\bibitem[{\citenamefont{Nakamuru et~al.}(1997)\citenamefont{Nakamuru, Matsuda,
  and Iwasa}}]{Nakamuru:1997}
\bibinfo{author}{\bibfnamefont{M.}~\bibnamefont{Nakamuru}},
  \bibinfo{author}{\bibfnamefont{H.}~\bibnamefont{Matsuda}}, \bibnamefont{and}
  \bibinfo{author}{\bibfnamefont{Y.}~\bibnamefont{Iwasa}}, \bibinfo{journal}{J.
  Theor. Biol.} \textbf{\bibinfo{volume}{184}}, \bibinfo{pages}{65}
  (\bibinfo{year}{1997}).

\bibitem[{\citenamefont{Langer et~al.}(2008)\citenamefont{Langer, Nowak, and
  Hauert}}]{Langer:2008}
\bibinfo{author}{\bibfnamefont{P.}~\bibnamefont{Langer}},
  \bibinfo{author}{\bibfnamefont{M.~A.} \bibnamefont{Nowak}}, \bibnamefont{and}
  \bibinfo{author}{\bibfnamefont{C.}~\bibnamefont{Hauert}},
  \bibinfo{journal}{Journal of Theoretical Biology}
  \textbf{\bibinfo{volume}{250}}, \bibinfo{pages}{634} (\bibinfo{year}{2008}),
  ISSN \bibinfo{issn}{00225193}.

\bibitem[{\citenamefont{Szab\'{o} and Szolnoki}(2009)}]{Szabo:2009}
\bibinfo{author}{\bibfnamefont{G.}~\bibnamefont{Szab\'{o}}} \bibnamefont{and}
  \bibinfo{author}{\bibfnamefont{A.}~\bibnamefont{Szolnoki}},
  \bibinfo{journal}{Phys. Rev. E} \textbf{\bibinfo{volume}{79}},
  \bibinfo{pages}{016106} (\bibinfo{year}{2009}).

\bibitem[{\citenamefont{Helbing and Yu}(2009)}]{Helbing:2009}
\bibinfo{author}{\bibfnamefont{D.}~\bibnamefont{Helbing}} \bibnamefont{and}
  \bibinfo{author}{\bibfnamefont{W.}~\bibnamefont{Yu}}, \bibinfo{journal}{Proc.
  Natl. Acad. Sci. USA} \textbf{\bibinfo{volume}{106}}, \bibinfo{pages}{3680}
  (\bibinfo{year}{2009}).

\bibitem[{\citenamefont{Fu et~al.}(2010)\citenamefont{Fu, Nowak, and
  Hauert}}]{Fu:2010}
\bibinfo{author}{\bibfnamefont{F.}~\bibnamefont{Fu}},
  \bibinfo{author}{\bibfnamefont{M.~A.} \bibnamefont{Nowak}}, \bibnamefont{and}
  \bibinfo{author}{\bibfnamefont{C.}~\bibnamefont{Hauert}},
  \bibinfo{journal}{Journal of Theoretical Biology}
  \textbf{\bibinfo{volume}{266}}, \bibinfo{pages}{358 } (\bibinfo{year}{2010}),
  ISSN \bibinfo{issn}{0022-5193}.

\bibitem[{\citenamefont{Szolnoki
  et~al.}(2009{\natexlab{a}})\citenamefont{Szolnoki, Vukov, and
  Szab\'o}}]{Szolnoki:2009}
\bibinfo{author}{\bibfnamefont{A.}~\bibnamefont{Szolnoki}},
  \bibinfo{author}{\bibfnamefont{J.}~\bibnamefont{Vukov}}, \bibnamefont{and}
  \bibinfo{author}{\bibfnamefont{G.}~\bibnamefont{Szab\'o}},
  \bibinfo{journal}{Phys. Rev. E} \textbf{\bibinfo{volume}{80}},
  \bibinfo{pages}{056112} (\bibinfo{year}{2009}{\natexlab{a}}).

\bibitem[{\citenamefont{Liu et~al.}(2012)\citenamefont{Liu, Jia, Zhang, and
  Wang}}]{Liu:2012}
\bibinfo{author}{\bibfnamefont{R.-R.} \bibnamefont{Liu}},
  \bibinfo{author}{\bibfnamefont{C.-X.} \bibnamefont{Jia}},
  \bibinfo{author}{\bibfnamefont{J.}~\bibnamefont{Zhang}}, \bibnamefont{and}
  \bibinfo{author}{\bibfnamefont{B.-H.} \bibnamefont{Wang}},
  \bibinfo{journal}{Physica A: Statistical Mechanics and its Applications}
  \textbf{\bibinfo{volume}{391}}, \bibinfo{pages}{4325 }
  (\bibinfo{year}{2012}), ISSN \bibinfo{issn}{0378-4371}.

\bibitem[{\citenamefont{Szolnoki
  et~al.}(2009{\natexlab{b}})\citenamefont{Szolnoki, Perc, Szab\'o, and
  Stark}}]{Szolnoki:2009a}
\bibinfo{author}{\bibfnamefont{A.}~\bibnamefont{Szolnoki}},
  \bibinfo{author}{\bibfnamefont{M.} \bibnamefont{Perc}},
  \bibinfo{author}{\bibfnamefont{G.}~\bibnamefont{Szab\'o}}, \bibnamefont{and}
  \bibinfo{author}{\bibfnamefont{H.-U.} \bibnamefont{Stark}},
  \bibinfo{journal}{Phys. Rev. E} \textbf{\bibinfo{volume}{80}},
  \bibinfo{pages}{021901} (\bibinfo{year}{2009}{\natexlab{b}}).

\bibitem[{\citenamefont{Alonso-Sanz}(2009)}]{AlonsoSanz:2009}
\bibinfo{author}{\bibfnamefont{R.}~\bibnamefont{Alonso-Sanz}},
  \bibinfo{journal}{Biosystems} \textbf{\bibinfo{volume}{97}},
  \bibinfo{pages}{90 } (\bibinfo{year}{2009}).

\bibitem[{\citenamefont{Wang et~al.}(2012)\citenamefont{Wang, Szolnoki, and
  Perc}}]{Wang:2012}
\bibinfo{author}{\bibfnamefont{Z.}~\bibnamefont{Wang}},
  \bibinfo{author}{\bibfnamefont{A.}~\bibnamefont{Szolnoki}}, \bibnamefont{and}
  \bibinfo{author}{\bibfnamefont{M.} \bibnamefont{Perc}},
  \bibinfo{journal}{Phys. Rev. E} \textbf{\bibinfo{volume}{85}},
  \bibinfo{pages}{037101} (\bibinfo{year}{2012}).

\bibitem[{\citenamefont{Ohtsuki et~al.}(2006)\citenamefont{Ohtsuki, Hauert,
  Lieberman, and Nowak}}]{Ohtsuki:2006}
\bibinfo{author}{\bibfnamefont{H.}~\bibnamefont{Ohtsuki}},
  \bibinfo{author}{\bibfnamefont{C.}~\bibnamefont{Hauert}},
  \bibinfo{author}{\bibfnamefont{E.}~\bibnamefont{Lieberman}},
  \bibnamefont{and} \bibinfo{author}{\bibfnamefont{M.~A.} \bibnamefont{Nowak}},
  \bibinfo{journal}{Natur} \textbf{\bibinfo{volume}{441}}, \bibinfo{pages}{502}
  (\bibinfo{year}{2006}).

\bibitem[{\citenamefont{van Baalen and Rand}(1998)}]{vanBaalen:1998}
\bibinfo{author}{\bibfnamefont{M.}~\bibnamefont{van Baalen}} \bibnamefont{and}
  \bibinfo{author}{\bibfnamefont{D.~A.} \bibnamefont{Rand}},
  \bibinfo{journal}{Journal of Theoretical Biology}
  \textbf{\bibinfo{volume}{193}}, \bibinfo{pages}{631 } (\bibinfo{year}{1998}).

\bibitem[{\citenamefont{Ifti et~al.}(2004)\citenamefont{Ifti, Killingback, and
  Doebeli}}]{Ifti:2004}
\bibinfo{author}{\bibfnamefont{M.}~\bibnamefont{Ifti}},
  \bibinfo{author}{\bibfnamefont{T.}~\bibnamefont{Killingback}},
  \bibnamefont{and} \bibinfo{author}{\bibfnamefont{M.}~\bibnamefont{Doebeli}},
  \bibinfo{journal}{J. Theor. Biol.} \textbf{\bibinfo{volume}{231}},
  \bibinfo{pages}{97 } (\bibinfo{year}{2004}), ISSN \bibinfo{issn}{0022-5193}.

\bibitem[{\citenamefont{Santos and Pacheco}(2005)}]{Santos:2005}
\bibinfo{author}{\bibfnamefont{F.~C.} \bibnamefont{Santos}} \bibnamefont{and}
  \bibinfo{author}{\bibfnamefont{J.~M.} \bibnamefont{Pacheco}},
  \bibinfo{journal}{Phys. Rev. Lett.} \textbf{\bibinfo{volume}{95}},
  \bibinfo{pages}{098104} (\bibinfo{year}{2005}).

\bibitem[{\citenamefont{Abramson and Kuperman}(2001)}]{Abramson:2001}
\bibinfo{author}{\bibfnamefont{G.}~\bibnamefont{Abramson}} \bibnamefont{and}
  \bibinfo{author}{\bibfnamefont{M.}~\bibnamefont{Kuperman}},
  \bibinfo{journal}{Phys. Rev. E} \textbf{\bibinfo{volume}{63}},
  \bibinfo{pages}{030901} (\bibinfo{year}{2001}).

\bibitem[{\citenamefont{Lieberman et~al.}(2005)\citenamefont{Lieberman, Hauert,
  and Nowak}}]{Lieberman:2005}
\bibinfo{author}{\bibfnamefont{E.}~\bibnamefont{Lieberman}},
  \bibinfo{author}{\bibfnamefont{C.}~\bibnamefont{Hauert}}, \bibnamefont{and}
  \bibinfo{author}{\bibfnamefont{M.~A.} \bibnamefont{Nowak}},
  \bibinfo{journal}{Nature} \textbf{\bibinfo{volume}{433}},
  \bibinfo{pages}{312} (\bibinfo{year}{2005}), ISSN \bibinfo{issn}{0028-0836}.

\bibitem[{\citenamefont{Szab\'o and F\'ath}(2007)}]{Szabo}
\bibinfo{author}{\bibfnamefont{G.}~\bibnamefont{Szab\'o}} \bibnamefont{and}
  \bibinfo{author}{\bibfnamefont{G.}~\bibnamefont{F\'ath}},
  \bibinfo{journal}{Phys. Rep.} \textbf{\bibinfo{volume}{446}},
  \bibinfo{pages}{97} (\bibinfo{year}{2007}).

\bibitem[{\citenamefont{Vukov et~al.}(2006)\citenamefont{Vukov, Szab\'o, and
  Szolnoki}}]{Vukov:2006}
\bibinfo{author}{\bibfnamefont{J.}~\bibnamefont{Vukov}},
  \bibinfo{author}{\bibfnamefont{G.}~\bibnamefont{Szab\'o}}, \bibnamefont{and}
  \bibinfo{author}{\bibfnamefont{A.}~\bibnamefont{Szolnoki}},
  \bibinfo{journal}{Phys. Rev. E} \textbf{\bibinfo{volume}{73}},
  \bibinfo{pages}{067103} (\bibinfo{year}{2006}).

\bibitem[{\citenamefont{Le~Gac and Doebeli}(2010)}]{LeGac:2009}
\bibinfo{author}{\bibfnamefont{M.}~\bibnamefont{Le~Gac}} \bibnamefont{and}
  \bibinfo{author}{\bibfnamefont{M.}~\bibnamefont{Doebeli}},
  \bibinfo{journal}{Evolution} \textbf{\bibinfo{volume}{64}},
  \bibinfo{pages}{522} (\bibinfo{year}{2010}), ISSN \bibinfo{issn}{1558-5646}.

\bibitem[{\citenamefont{Szab\'o and T\"oke}(1998)}]{Szabo:1998}
\bibinfo{author}{\bibfnamefont{G.}~\bibnamefont{Szab\'o}} \bibnamefont{and}
  \bibinfo{author}{\bibfnamefont{C.}~\bibnamefont{T\"oke}},
  \bibinfo{journal}{Phys. Rev. E} \textbf{\bibinfo{volume}{58}},
  \bibinfo{pages}{69} (\bibinfo{year}{1998}).

\bibitem[{\citenamefont{Szab\'o and Hauert}(2002)}]{Szab'o:2002}
\bibinfo{author}{\bibfnamefont{G.}~\bibnamefont{Szab\'o}} \bibnamefont{and}
  \bibinfo{author}{\bibfnamefont{C.}~\bibnamefont{Hauert}},
  \bibinfo{journal}{Phys. Rev. Lett.} \textbf{\bibinfo{volume}{89}},
  \bibinfo{pages}{118101} (\bibinfo{year}{2002}).

\bibitem[{\citenamefont{Szab\'o et~al.}(2005)\citenamefont{Szab\'o, Vukov, and
  Szolnoki}}]{Szabo:2005}
\bibinfo{author}{\bibfnamefont{G.}~\bibnamefont{Szab\'o}},
  \bibinfo{author}{\bibfnamefont{J.}~\bibnamefont{Vukov}}, \bibnamefont{and}
  \bibinfo{author}{\bibfnamefont{A.}~\bibnamefont{Szolnoki}},
  \bibinfo{journal}{Phys. Rev. E} \textbf{\bibinfo{volume}{72}},
  \bibinfo{eid}{047107} (pages~\bibinfo{numpages}{4}) (\bibinfo{year}{2005}).

\bibitem[{\citenamefont{Chiong and Kirley}(2012)}]{Chiong:2012}
\bibinfo{author}{\bibfnamefont{R.}~\bibnamefont{Chiong}} \bibnamefont{and}
  \bibinfo{author}{\bibfnamefont{M.}~\bibnamefont{Kirley}},
  \bibinfo{journal}{Physica A: Statistical Mechanics and its Applications}
  \textbf{\bibinfo{volume}{391}}, \bibinfo{pages}{3915 }
  (\bibinfo{year}{2012}).

\bibitem[{\citenamefont{Enquist and Leimar}(1993)}]{Enquist:1993}
\bibinfo{author}{\bibfnamefont{M.}~\bibnamefont{Enquist}} \bibnamefont{and}
  \bibinfo{author}{\bibfnamefont{O.}~\bibnamefont{Leimar}},
  \bibinfo{journal}{Anim. Behav.} \textbf{\bibinfo{volume}{45}},
  \bibinfo{pages}{747} (\bibinfo{year}{1993}).

\bibitem[{\citenamefont{Vainstein et~al.}(2007)\citenamefont{Vainstein, Silva,
  and Arenzon}}]{Vainstein:2007}
\bibinfo{author}{\bibfnamefont{M.~H.} \bibnamefont{Vainstein}},
  \bibinfo{author}{\bibfnamefont{A.~T.~C.} \bibnamefont{Silva}},
  \bibnamefont{and} \bibinfo{author}{\bibfnamefont{J.~J.}
  \bibnamefont{Arenzon}}, \bibinfo{journal}{J. Theor. Biol.}
  \textbf{\bibinfo{volume}{244}}, \bibinfo{pages}{722} (\bibinfo{year}{2007}).

\bibitem[{\citenamefont{Sicardi et~al.}(2009)\citenamefont{Sicardi, Fort,
  Vainstein, and Arenzon}}]{Sicardi:2009}
\bibinfo{author}{\bibfnamefont{E.~A.} \bibnamefont{Sicardi}},
  \bibinfo{author}{\bibfnamefont{H.}~\bibnamefont{Fort}},
  \bibinfo{author}{\bibfnamefont{M.~H.} \bibnamefont{Vainstein}},
  \bibnamefont{and} \bibinfo{author}{\bibfnamefont{J.~J.}
  \bibnamefont{Arenzon}}, \bibinfo{journal}{J. Theor. Biol.}
  \textbf{\bibinfo{volume}{256}}, \bibinfo{pages}{240} (\bibinfo{year}{2009}).

\bibitem[{\citenamefont{Helbing and Yu}(2008)}]{Helbing:2008}
\bibinfo{author}{\bibfnamefont{D.}~\bibnamefont{Helbing}} \bibnamefont{and}
  \bibinfo{author}{\bibfnamefont{W.}~\bibnamefont{Yu}}, \bibinfo{journal}{Adv.
  Compl. Syst.} \textbf{\bibinfo{volume}{11}}, \bibinfo{pages}{641}
  (\bibinfo{year}{2008}).

\bibitem[{\citenamefont{Diggle et~al.}(2007)\citenamefont{Diggle, Griffin,
  Campbell, and West}}]{Diggle}
\bibinfo{author}{\bibfnamefont{S.~P.} \bibnamefont{Diggle}},
  \bibinfo{author}{\bibfnamefont{A.~S.} \bibnamefont{Griffin}},
  \bibinfo{author}{\bibfnamefont{G.~S.} \bibnamefont{Campbell}},
  \bibnamefont{and} \bibinfo{author}{\bibfnamefont{S.~A.} \bibnamefont{West}},
  \bibinfo{journal}{Nature} \textbf{\bibinfo{volume}{450}},
  \bibinfo{pages}{411} (\bibinfo{year}{2007}).

\bibitem[{\citenamefont{Gillespie}(1976)}]{Gillespie}
\bibinfo{author}{\bibfnamefont{D.}~\bibnamefont{Gillespie}},
  \bibinfo{journal}{Journal of Computational Physics}
  \textbf{\bibinfo{volume}{22}}, \bibinfo{pages}{403} (\bibinfo{year}{1976}).

\bibitem[{\citenamefont{Nowak et~al.}(2004)\citenamefont{Nowak, Sasaki, Taylor,
  and Fudenberg}}]{Nowak:2004}
\bibinfo{author}{\bibfnamefont{M.~A.} \bibnamefont{Nowak}},
  \bibinfo{author}{\bibfnamefont{A.}~\bibnamefont{Sasaki}},
  \bibinfo{author}{\bibfnamefont{C.}~\bibnamefont{Taylor}}, \bibnamefont{and}
  \bibinfo{author}{\bibfnamefont{D.}~\bibnamefont{Fudenberg}},
  \bibinfo{journal}{Nature} \textbf{\bibinfo{volume}{428}},
  \bibinfo{pages}{646} (\bibinfo{year}{2004}).

\bibitem[{\citenamefont{Traulsen et~al.}(2005)\citenamefont{Traulsen, Claussen,
  and Hauert}}]{Traulsen:2005}
\bibinfo{author}{\bibfnamefont{A.}~\bibnamefont{Traulsen}},
  \bibinfo{author}{\bibfnamefont{J.~C.} \bibnamefont{Claussen}},
  \bibnamefont{and} \bibinfo{author}{\bibfnamefont{C.}~\bibnamefont{Hauert}},
  \bibinfo{journal}{Phys. Rev. Lett.} \textbf{\bibinfo{volume}{95}},
  \bibinfo{pages}{238701} (\bibinfo{year}{2005}).

\bibitem[{\citenamefont{Cremer et~al.}(2009)\citenamefont{Cremer, Reichenbach,
  and Frey}}]{Cremer:2009}
\bibinfo{author}{\bibfnamefont{J.}~\bibnamefont{Cremer}},
  \bibinfo{author}{\bibfnamefont{T.}~\bibnamefont{Reichenbach}},
  \bibnamefont{and} \bibinfo{author}{\bibfnamefont{E.}~\bibnamefont{Frey}},
  \bibinfo{journal}{New J. Phys.} \textbf{\bibinfo{volume}{11}},
  \bibinfo{pages}{093029} (\bibinfo{year}{2009}).

\bibitem[{\citenamefont{Galla}(2009)}]{Galla:2009}
\bibinfo{author}{\bibfnamefont{T.}~\bibnamefont{Galla}},
  \bibinfo{journal}{Phys. Rev. Lett.} \textbf{\bibinfo{volume}{103}},
  \bibinfo{pages}{198702} (\bibinfo{year}{2009}).

\bibitem[{\citenamefont{Cremer et~al.}(2011)\citenamefont{Cremer, Melbinger,
  and Frey}}]{CremerPRE:2011}
\bibinfo{author}{\bibfnamefont{J.}~\bibnamefont{Cremer}},
  \bibinfo{author}{\bibfnamefont{A.}~\bibnamefont{Melbinger}},
  \bibnamefont{and} \bibinfo{author}{\bibfnamefont{E.}~\bibnamefont{Frey}},
  \bibinfo{journal}{Phys. Rev. E} \textbf{\bibinfo{volume}{84}},
  \bibinfo{pages}{051921} (\bibinfo{year}{2011}).

\bibitem[{\citenamefont{Hinrichsen}(2000)}]{Hinrichsen:2000}
\bibinfo{author}{\bibfnamefont{H.}~\bibnamefont{Hinrichsen}},
  \bibinfo{journal}{Advances in Physics} \textbf{\bibinfo{volume}{49}},
  \bibinfo{pages}{815} (\bibinfo{year}{2000}).

\bibitem[{\citenamefont{Stauffer and Aharony}(1992)}]{Stauffer:1992}
\bibinfo{author}{\bibfnamefont{D.}~\bibnamefont{Stauffer}} \bibnamefont{and}
  \bibinfo{author}{\bibfnamefont{A.}~\bibnamefont{Aharony}},
  \emph{\bibinfo{title}{Introduction to Percolation Theory}}
  (\bibinfo{publisher}{Taylor \& Francis}, \bibinfo{year}{1992}).

\bibitem[{\citenamefont{Hinrichsen}(2006)}]{Hinrichsen:2006}
\bibinfo{author}{\bibfnamefont{H.}~\bibnamefont{Hinrichsen}},
  \bibinfo{journal}{Physica A} \textbf{\bibinfo{volume}{369}},
  \bibinfo{pages}{1} (\bibinfo{year}{2006}).

\bibitem[{\citenamefont{Dickman and Tretyakov}(1995)}]{Dickman:1995}
\bibinfo{author}{\bibfnamefont{R.}~\bibnamefont{Dickman}} \bibnamefont{and}
  \bibinfo{author}{\bibfnamefont{A.~Y.} \bibnamefont{Tretyakov}},
  \bibinfo{journal}{Phys. Rev. E} \textbf{\bibinfo{volume}{52}},
  \bibinfo{pages}{3218} (\bibinfo{year}{1995}).

\bibitem[{\citenamefont{Ben-Naim et~al.}(1996)\citenamefont{Ben-Naim,
  Frachebourg, and Krapivsky}}]{Ben-Naim:1995}
\bibinfo{author}{\bibfnamefont{E.}~\bibnamefont{Ben-Naim}},
  \bibinfo{author}{\bibfnamefont{L.}~\bibnamefont{Frachebourg}},
  \bibnamefont{and} \bibinfo{author}{\bibfnamefont{P.~L.}
  \bibnamefont{Krapivsky}}, \bibinfo{journal}{Phys. Rev. E}
  \textbf{\bibinfo{volume}{53}}, \bibinfo{pages}{3078} (\bibinfo{year}{1996}).

\bibitem[{\citenamefont{\'{O}dor}(2004)}]{Odor:2004}
\bibinfo{author}{\bibfnamefont{G.}~\bibnamefont{\'{O}dor}},
  \bibinfo{journal}{Rev. Mod. Phys.} \textbf{\bibinfo{volume}{76}},
  \bibinfo{pages}{663} (\bibinfo{year}{2004}).

\bibitem[{\citenamefont{Janssen}(1981)}]{Janssen:1981}
\bibinfo{author}{\bibfnamefont{H.~K.} \bibnamefont{Janssen}},
  \bibinfo{journal}{Z. Phys.} \textbf{\bibinfo{volume}{42}},
  \bibinfo{pages}{151} (\bibinfo{year}{1981}).

\bibitem[{\citenamefont{Grassberger}(1982)}]{Grassberger:1982}
\bibinfo{author}{\bibfnamefont{P.}~\bibnamefont{Grassberger}},
  \bibinfo{journal}{Z. Phys.} \textbf{\bibinfo{volume}{47}},
  \bibinfo{pages}{365} (\bibinfo{year}{1982}).

\bibitem[{\citenamefont{Qin et~al.}(2008)\citenamefont{Qin, Chen, Zhao, and
  Shi}}]{Qin:2008}
\bibinfo{author}{\bibfnamefont{S.-M.} \bibnamefont{Qin}},
  \bibinfo{author}{\bibfnamefont{Y.}~\bibnamefont{Chen}},
  \bibinfo{author}{\bibfnamefont{X.-Y.} \bibnamefont{Zhao}}, \bibnamefont{and}
  \bibinfo{author}{\bibfnamefont{J.}~\bibnamefont{Shi}},
  \bibinfo{journal}{Phys. Rev. E} \textbf{\bibinfo{volume}{78}},
  \bibinfo{pages}{041129} (\bibinfo{year}{2008}).

\bibitem[{\citenamefont{Fletcher and Doebeli}(2006)}]{Fletcher:2006}
\bibinfo{author}{\bibfnamefont{J.~A.} \bibnamefont{Fletcher}} \bibnamefont{and}
  \bibinfo{author}{\bibfnamefont{M.}~\bibnamefont{Doebeli}},
  \bibinfo{journal}{Journal of Evolutionary Biology}
  \textbf{\bibinfo{volume}{19}}, \bibinfo{pages}{1389} (\bibinfo{year}{2006}),
  ISSN \bibinfo{issn}{1420-9101}.

\bibitem[{\citenamefont{Rainey and Rainey}(2004)}]{Rainey:2003}
\bibinfo{author}{\bibfnamefont{P.~B.} \bibnamefont{Rainey}} \bibnamefont{and}
  \bibinfo{author}{\bibfnamefont{K.}~\bibnamefont{Rainey}},
  \bibinfo{journal}{Nature} \textbf{\bibinfo{volume}{425}}, \bibinfo{pages}{72}
  (\bibinfo{year}{2004}).

\bibitem[{\citenamefont{Cremer et~al.}(2012)\citenamefont{Cremer, Melbinger,
  and Frey}}]{Cremer:2012}
\bibinfo{author}{\bibfnamefont{J.}~\bibnamefont{Cremer}},
  \bibinfo{author}{\bibfnamefont{A.}~\bibnamefont{Melbinger}},
  \bibnamefont{and} \bibinfo{author}{\bibfnamefont{E.}~\bibnamefont{Frey}},
  \bibinfo{journal}{Scientific Reports} \textbf{\bibinfo{volume}{2}}
  (\bibinfo{year}{2012}), ISSN \bibinfo{issn}{2045-2322}.

\bibitem[{\citenamefont{Melbinger et~al.}(2010)\citenamefont{Melbinger, Cremer,
  and Frey}}]{Melbinger:2010}
\bibinfo{author}{\bibfnamefont{A.}~\bibnamefont{Melbinger}},
  \bibinfo{author}{\bibfnamefont{J.}~\bibnamefont{Cremer}}, \bibnamefont{and}
  \bibinfo{author}{\bibfnamefont{E.}~\bibnamefont{Frey}},
  \bibinfo{journal}{Phys. Rev. Lett.} \textbf{\bibinfo{volume}{105}},
  \bibinfo{pages}{178101} (\bibinfo{year}{2010}).

\bibitem[{\citenamefont{West et~al.}(2007)\citenamefont{West, Griffin, and
  Gardner}}]{West:2007b}
\bibinfo{author}{\bibfnamefont{S.~A.} \bibnamefont{West}},
  \bibinfo{author}{\bibfnamefont{A.~S.} \bibnamefont{Griffin}},
  \bibnamefont{and} \bibinfo{author}{\bibfnamefont{A.}~\bibnamefont{Gardner}},
  \bibinfo{journal}{Curr. Biol.} \textbf{\bibinfo{volume}{17}},
  \bibinfo{pages}{R661} (\bibinfo{year}{2007}).

\bibitem[{\citenamefont{Hallatschek et~al.}(2007)\citenamefont{Hallatschek,
  Hersen, Ramanathan, and Nelson}}]{Hallatschek:2007}
\bibinfo{author}{\bibfnamefont{O.}~\bibnamefont{Hallatschek}},
  \bibinfo{author}{\bibfnamefont{P.}~\bibnamefont{Hersen}},
  \bibinfo{author}{\bibfnamefont{S.}~\bibnamefont{Ramanathan}},
  \bibnamefont{and} \bibinfo{author}{\bibfnamefont{D.~R.}
  \bibnamefont{Nelson}}, \bibinfo{journal}{Proceedings of the National Academy
  of Sciences} \textbf{\bibinfo{volume}{104}}, \bibinfo{pages}{19926}
  (\bibinfo{year}{2007}).

\bibitem[{\citenamefont{Hallatschek and Nelson}(2008)}]{Hallatschek2008158}
\bibinfo{author}{\bibfnamefont{O.}~\bibnamefont{Hallatschek}} \bibnamefont{and}
  \bibinfo{author}{\bibfnamefont{D.~R.} \bibnamefont{Nelson}},
  \bibinfo{journal}{Theoretical Population Biology}
  \textbf{\bibinfo{volume}{73}}, \bibinfo{pages}{158 } (\bibinfo{year}{2008}).

\bibitem[{\citenamefont{Kuhr et~al.}(2011)\citenamefont{Kuhr, Leisner, and
  Frey}}]{Kuhr:2011}
\bibinfo{author}{\bibfnamefont{J.-T.} \bibnamefont{Kuhr}},
  \bibinfo{author}{\bibfnamefont{M.}~\bibnamefont{Leisner}}, \bibnamefont{and}
  \bibinfo{author}{\bibfnamefont{E.}~\bibnamefont{Frey}}, \bibinfo{journal}{New
  J. Phys.} \textbf{\bibinfo{volume}{13}}, \bibinfo{pages}{113013}
  (\bibinfo{year}{2011}).

\bibitem[{\citenamefont{Korolev and Nelson}(2011)}]{Korolev2011}
\bibinfo{author}{\bibfnamefont{K.~S.} \bibnamefont{Korolev}} \bibnamefont{and}
  \bibinfo{author}{\bibfnamefont{D.~R.} \bibnamefont{Nelson}},
  \bibinfo{journal}{Phys. Rev. Let.} \textbf{\bibinfo{volume}{107}},
  \bibinfo{pages}{088103} (\bibinfo{year}{2011}).

\bibitem[{\citenamefont{Lavrentovich et~al.}(2013)\citenamefont{Lavrentovich,
  Korolev, and Nelson}}]{Lavrentovich2013}
\bibinfo{author}{\bibfnamefont{M.~O.} \bibnamefont{Lavrentovich}},
  \bibinfo{author}{\bibfnamefont{K.~S.} \bibnamefont{Korolev}},
  \bibnamefont{and} \bibinfo{author}{\bibfnamefont{D.~R.}
  \bibnamefont{Nelson}}, \bibinfo{journal}{to be published}
  (\bibinfo{year}{2013}).

\end{thebibliography}

\end{document}